\documentclass[%
preprint,
showpacs,
 amsmath,
 amssymb,
nobalancelastpage]{revtex4-1}

\usepackage{color}
\usepackage{graphicx}% Include figure files
\usepackage{dcolumn}% Align table columns on decimal point
\usepackage{bm}% bold math
\usepackage{esint}
\usepackage{todo}
\usepackage[colorlinks,allcolors=black]{hyperref}
\usepackage[capitalise,nosort]{cleveref} %for smart references, \cref
\usepackage{csquotes}
\usepackage{verbatim}
\usepackage{hhline}
\usepackage[normalem]{ulem}

\usepackage[english]{babel} 
\usepackage{microtype}
\usepackage{siunitx}
\usepackage{dsfont}
\usepackage[dvipsnames]{xcolor}

\usepackage{eucal}

{}

\newcommand{\q}[1]{``#1"}

\newcommand{\dpol}{\ensuremath{{\boldsymbol{\phi}}}}

\crefname{appsec}{appendix}{appendices}

\begin{document}

\title{Temperature transitions and degeneracy in the control of small clusters with a macroscopic field}

\author{Francesco Boccardo}
\email{francesco.boccardo@univ-lyon1.fr}
\author{Olivier Pierre-Louis}
\email{olivier.pierre-louis@univ-lyon1.fr}
\affiliation{Institut Lumi\`ere Mati\`ere, UMR5306 Universit\'e Lyon 1 - CNRS, 69622 Villeurbanne, France}
\date{\today}
\begin{abstract}
We present a numerical investigation of the control of 
few-particle fluctuating clusters with a macroscopic field.
Our goal is to reach a  given target cluster shape is minimum time.
This question is formulated as a first passage problem in the 
space of cluster configurations.
We find the optimal policy to set the macroscopic field
as a function of the observed shape using dynamic programming.
Our results show that the optimal policy is non-unique, and its
degeneracy is mainly related to symmetries shared by the
initial shape, the force and the target shape.
The total fraction of shapes for which optimal choice of the force is non-unique 
vanishes as the cluster size increases.
Furthermore, the optimal policy exhibits a discrete set of transitions when the temperature
is varied. Each transition leads to a discontinuity in the derivative
of the time to reach with target with respect to temperature.
As the size of the cluster increases, the 
change in the policy due to temperature transitions 
grows like the total number of configurations and a continuum limit emerges.
\end{abstract}

\maketitle
%  \tableofcontents
%%%%%%%%%%%%%%%%%%%%%%%%%%%%%%%%%%%%%%%%%%%%%%%%%%
%%%%%%%%%%%%%%%%%%%%%%%%%%%%%%%%%%%%%%%%%%%%%%%%%%
%%%%%%%%%%%%%%%%%%%%%%%%%%%%%%%%%%%%%%%%%%%%%%%%%%

%%%%%%%%%%%%%%%%%%%%%%%%%%%%%%%%%%%%%%%%%%%%%%%%%%
%%%%%%%%%%%%%%%%%%%%%%%%%%%%%%%%%%%%%%%%%%%%%%%%%%
%%%%%%%%%%%%%%%%%%%%%%%%%%%%%%%%%%%%%%%%%%%%%%%%%%
% \clearpage
%%%%%%%%%%%%%%%%%%%%%%%%%%%%%%%%%%%%%%%%%%%%%%%%%%
\section{Introduction}
%%%%%%%%%%%%%%%%%%%%%%%%%%%%%%%%%%%%%%%%%%%%%%%%%%

Forces produced by macroscopic electric fields or macroscopic thermal gradients
have been successfully used to displace single nano-scale objects, but also clusters of interacting 
nano-objects ranging from atoms~\cite{Tao2010,Curiotto2019} to fullerenes~\cite{Srivastava2009}
or microbial aggregates~\cite{Schneiderheinze2000}.
However, these forces also have the ability to alter the shape of the clusters. 
For example in the case of electromigration of large monolayer clusters,
complex cluster shapes are known to emerge from instabilities~\cite{Kumar2017,PierreLouis2000} 
and from the coupling
of the electric field direction to the cluster edge anisotropy~\cite{Kuhn2005,Kauttonen2012,Curiotto2019}.
Another example of non-trivial morphologies
emerging in the presence of an external field
is the shaping of small nano-particle clusters by light~\cite{McCormack2018}.
In these systems, efforts have been devoted to the identification of the
non-trivial set of morphologies that emerge from 
cluster dynamics under external fields~\cite{Kumar2017,PierreLouis2000,Kuhn2005,Kauttonen2012,McCormack2018}. 
Here, we wish to ask a different question: 
can we start by defining arbitrarily a target cluster shape that we want to reach,
and then find a way to reach it in a given physical system?
We are more specifically focusing on the case of two-dimensional clusters
that undergo spontaneous reshaping due to thermal fluctuations~\cite{Khare1995}
that are ubiquitous at the nanoscale.
We wish to use the external force to bias the fluctuation-induced cluster shape exploration
in order to reach a predefined target shape.
Achieving this goal would open novel perspectives in the design of 
nanostructures and colloid clusters.

Our approach is to set the value
of a external macroscopic field as a function of the observed cluster configuration
so as to minimize the time to reach a given target configuration. 
A choice of the field for each possible configuration is called a policy.
We therefore address the question of finding the policy that minimizes
the expected time to reach a given target cluster configuration.
We formulate this problem as the optimization
of first passage times on a graph (the vertices of this graph
are the configurations of the cluster, and the edges are the 
physical transitions). This optimization problem is 
solved using dynamic programming~\cite{Sutton1998,Bellman1957}.
We exemplify this approach on the specific case of a migration force 
that can take 3 values along a fixed direction (positive, negative, and zero),
and a cluster that changes shape via particle hopping along the periphery.

In a previous paper~\cite{Boccardo2022a}, we have shown that 
this approach allows one to drive few-particle clusters
towards a given target shape using a macroscopic external field.
These results showed that the macroscopic field can allow one to gain
orders of magnitude in the time to reach the target shape, and that
such strategies should apply quantitatively to colloid clusters.
In the present paper, we identify two fundamental
properties of the policy: degeneracy and temperature transitions.
Degeneracy is defined as the non-uniqueness of the optimal policy.
Degeneracy is shown to blow up exponentially when increasing cluster size.
However, the fraction of cluster configurations for which the choice
of the force is degenerate vanishes for large clusters.
In addition, degeneracy is shown to be mainly dictated by cluster symmetries that are compatible with those
of the external field. We also identify
temperature transitions, i.e. changes in the optimal policy
that occur when varying the temperature. 
These transitions lead to discontinuities in the 
derivative of the minimal time to reach the target with respect to temperature.
When the size of the cluster increases, the number of transitions increases.
For large-enough clusters, the density of change in the policy due to the temperature transitions
tends to a smooth function of the temperature. We conclude that the number
of states where the policy changes when varying the temperature
is proportional to the total number of states.

%%%%%%%%%%%%%%%%%%%%%%%%%%%%%%%%%%%%%%%%%%%%%%%%%%
\section{Model}
%%%%%%%%%%%%%%%%%%%%%%%%%%%%%%%%%%%%%%%%%%%%%%%%%%
\label{s:model}

We start with the presentation of the model, which is identical to that of 
our previous paper~\cite{Boccardo2022a}.
We consider a bond-breaking model on a square lattice with lattice parameter $a$. 
The nearest-neighbor bond energy is denoted as $J$. 
Each cluster configuration on the lattice defines a state $s$ of the cluster.
Two states are identical if one state can be obtained from the other
with the help of lattice translations.

\begin{figure}[ht]
    \includegraphics[width=.5\linewidth]{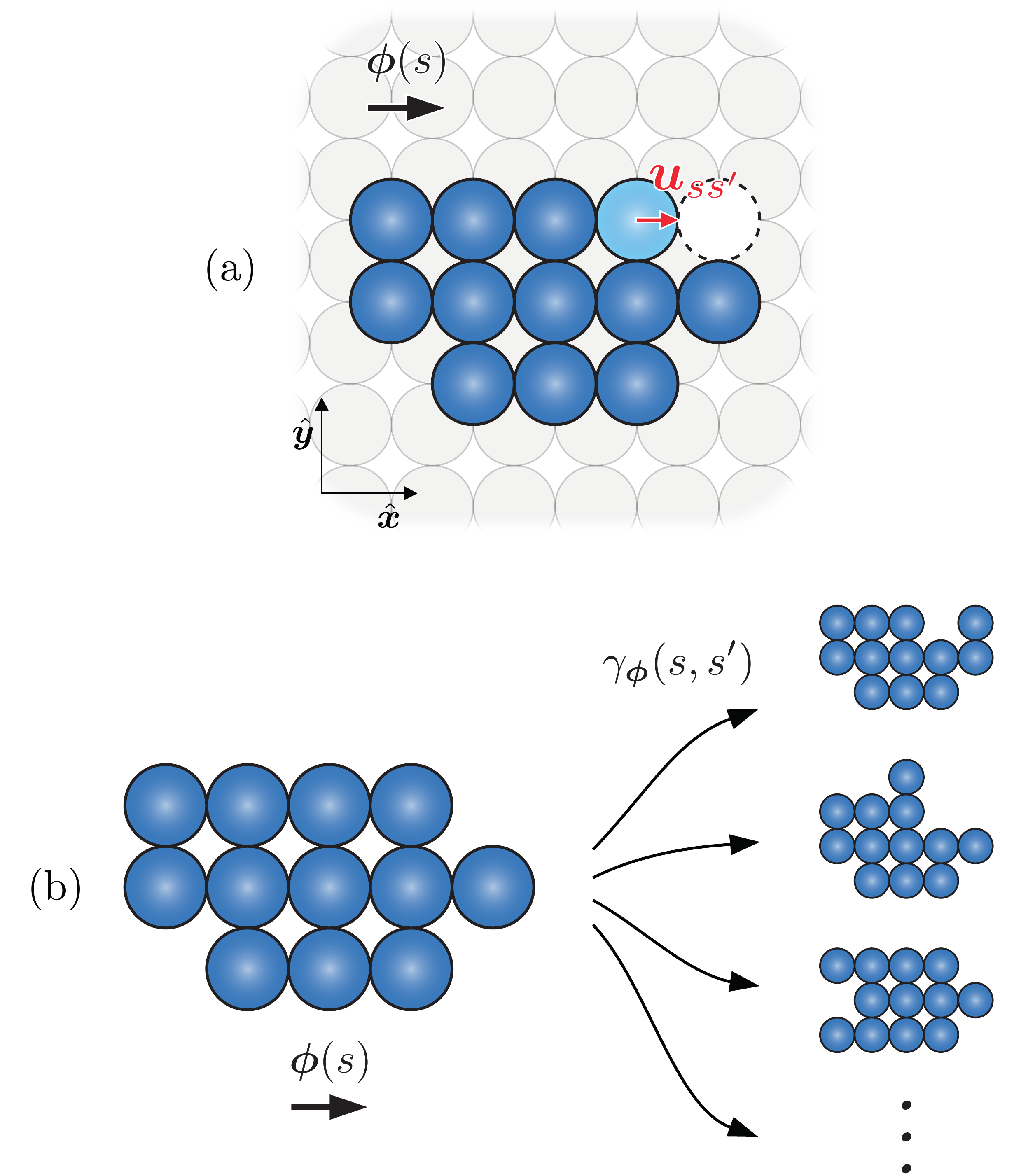}
    \caption{Lattice model for biased edge diffusion. 
    (a) Hopping of a particle. In this example, $\bm{u}_{ss'}=+(1/2)\hat{\bm{x}}$ and $n_{ss'}=2$.
    (b) For a given configuration $s$ and force $\dpol(s)$, 
    the hopping rates $\gamma_{\dpol}(s,s')$ to reachable states $s'$ are given by \cref{eq:hopping_rate}.
    }
    \label{fig:lattice_model}
\end{figure}

The state $s$ of the cluster can change to another state $s'$ via the motion
of a single particle to its nearest or next-nearest neighbor sites.
Moves are allowed only if they do not break the cluster.
As shown in \cref{fig:lattice_model}(a), this leads to edge diffusion dynamics.
In order to reduce computational costs, we discard moves of particles that have initially 4 nearest neighbors.
Some examples of possible moves are shown in \cref{fig:lattice_model}(b).

We assume that an external force $\bm{F}$ biases diffusion.
This bias is implemented in the model following the same lines as in 
previous kinetic Monte Carlo studies of monolayer clusters
in the presence of electromigration~\cite{PierreLouis2000,Liu1998}. 
The function $\dpol(s)$, called policy, 
sets a value of the force in each state $s$, so that $\bm{F}={\boldsymbol \phi}(s)$.
We choose a force that can take three different values
$\bm{F}=F_0\hat{\bm{x}}$, $\bm 0$, or $F_0 \hat{\bm{x}}$, with $F_0>0$
and $\hat{\bm{x}}$ the unit vector along the [10] Miller direction of the square lattice,
as shown in \cref{fig:lattice_model}(a).

The particle hopping rate along the edge of the cluster reads
\begin{align}
    \gamma_{\dpol}(s,s') = \nu \exp[-E_{\dpol}(s,s')/k_BT]\,,
    \label{eq:hopping_rate}
\end{align}
where $ k_B T $ is the thermal energy and $\nu$ is an attempt frequency.
The hopping barrier is composed of two contributions~\cite{Liu1998,PierreLouis2000}
\begin{equation}\label{eq:energy_barrier}
E_{\dpol}(s,s') = n_{ss'}J - \dpol(s)\cdot\bm{u}_{ss'} \, .
\end{equation}
The first contribution accounts for the number $n_{ss'}$ 
of nearest neighbor bonds before hopping
(in state $s$) that need to be broken for the move to occur.
The energy cost for breaking each bond is denoted as $J$.
This contribution is compatible with detailed balance
in the absence of forces~\cite{Boccardo2022b}. Hence, the system can reach thermodynamic equilibrium
in the absence of force, with an edge energy which is identical
to that of the low-temperature Ising model~\cite{Rottman1981,Krishnamachari1996,Khare1995,Saito1996}.
The vector $\bm{u}_{ss'}$ is the vector between the equilibrium position of the particle
before the move and the saddle point of the diffusion potential~\cite{Kandel1996,PierreLouis2000}.
For definiteness, we assume $\bm{u}_{ss'}$ is halfway between
the initial and the final positions of the particle, as depicted in \cref{fig:lattice_model}(a).
We also define the scalar $u_{ss'}=\hat{\bm{x}}\cdot\bm{u}_{ss'}$, 
so that $u_{ss'}$ takes one of the three values $-1/2,0,1/2$
depending on the relative directions of the force and of the move.

As an additional remark, for the description of particle hopping by means of transition-state
theory to make sense, the energy barrier has to be (i) positive, and (ii) larger than the
thermal energy $k_BT$~\cite{VanKampen1992}.
The first condition (i) requires that $\dpol(s)\cdot\bm{u}_{ss'}<J$ for any $\bm{u}_{ss'}$,
leading to the condition $F_0/J<2$.
The second condition (ii)  imposes that
the temperature has to be low enough $k_BT<J$.
This latter condition is actually not an exact inequality,
and should rather be considered as an approximate constraint,
which indicates that an Arrhenius form of the rates can be derived
as an asymptotic limit when $k_BT/J\ll1$,
but the expression is expected to be qualitatively correct
up to $k_BT/J\sim 1$.

In the following, we use normalized
variables with $a=1$, $J=1$, $\nu=1$ and $k_B=1$.

%%%%%%%%%%%%%%%%%%%%%%%%%%%%%%%%%%%%%%%%%%%%%%%%%%
\section{First passage times}
%%%%%%%%%%%%%%%%%%%%%%%%%%%%%%%%%%%%%%%%%%%%%%%%%%

\begin{figure}[ht]
	\includegraphics[width=.5\linewidth]{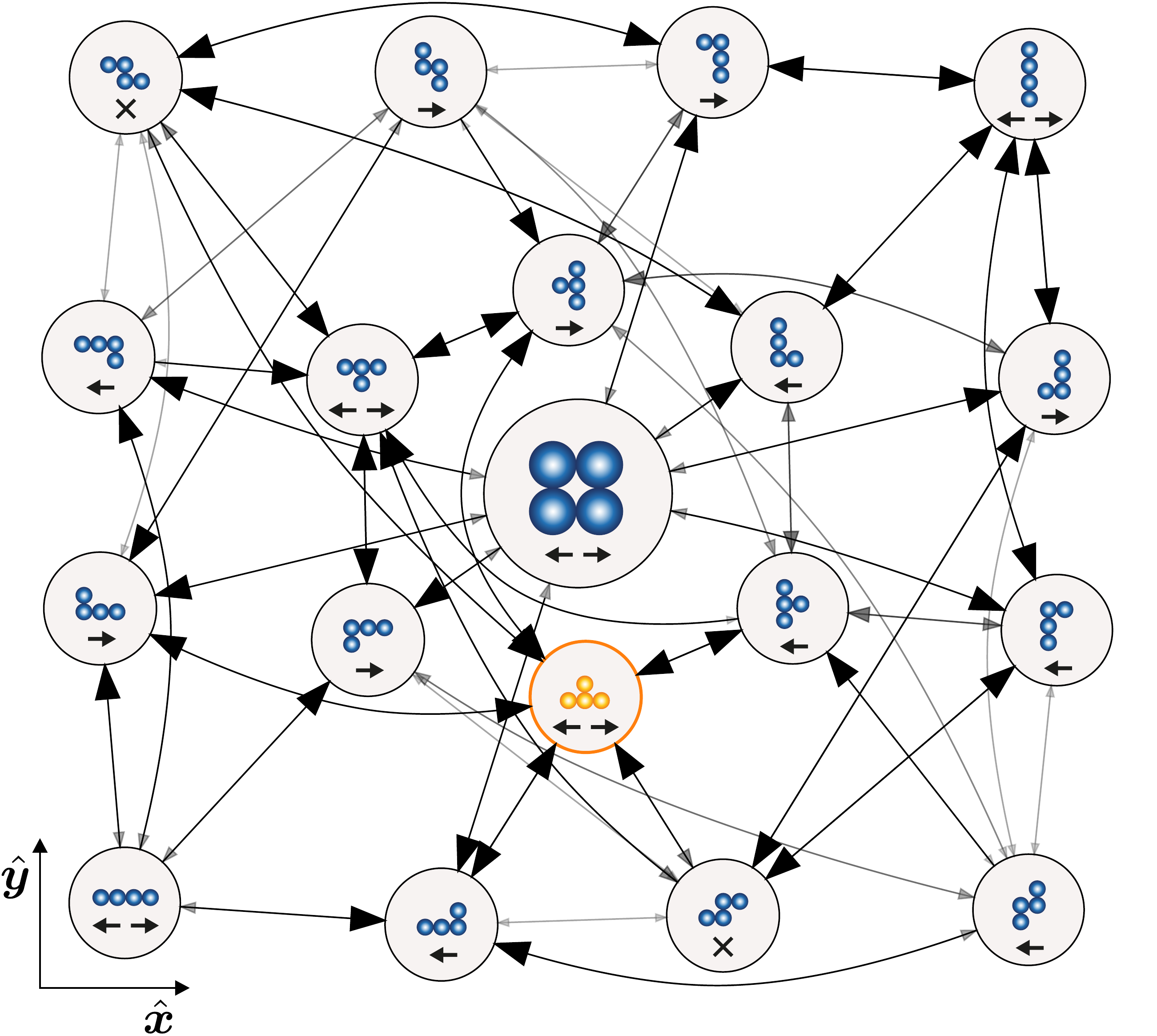}
	\caption{Graph of configurations for a tetramer cluster ($N=4$) at $T=0.24$.
	The node size is proportional to the expected residence time $t_{\boldsymbol \phi}(s)$.
	The thickness and shade of the edges are proportional to the transition probability $p_{\boldsymbol \phi}(s,s')$. 
	Arrows in the nodes represent an optimal policy ${\boldsymbol \phi}_*(s)$ to reach the orange target shape (crosses correspond to a zero force). Note that, for this target, there are several states in which the force can be set equivalently to the right or left (degenerate optimal action). }
	\label{fig:graph_dynamics_tetramer}
\end{figure}

The stochastic dynamics of the cluster can be seen 
as a random walk in the space of configurations.
This space can be represented as a graph.
The vertices of the graph are the states of the clusters,
and the edges correspond to the possible moves from one state to another.
An example of graph is shown in \cref{fig:graph_dynamics_tetramer}.

Under a given policy $\dpol$ which sets the value of the force in each state, 
the expected residence time in state $s$ is
\begin{align}
\label{eq:mean_res_time}
t_{\dpol}(s)&= \dfrac{1}{\sum_{s'\in{\mathcal{B}}_s}\gamma_\dpol(s,s')} \,.
\end{align}
The sum over $s'$ is taken over the set ${\mathcal{B}}_s$ of neighboring states that can be reached from $s$ by 
a single one-particle move.
Furthermore, the probability of transition from $s$ to a state $s'\in {\mathcal{B}}_s$ reads
\begin{align}
\label{eq:transition_proba}
p_\dpol(s,s') &= \gamma_\dpol(s,s')t_\dpol(s) \,.
\end{align}
Due to the Markovian character of the dynamics, 
the expected first passage time  $\tau_\dpol(s,\bar{s})$ from state $s$ 
to the target state $\bar{s}$  is equal to the residence time $t_\dpol(s)$
plus the first passage time in the neighboring states $s'$ after the move,
weighted by the probability $p_{\dpol}(s,s')$ 
to go to these states.
Hence~\cite{VanKampen1992,Boccardo2022a}
\begin{align}
\label{eq:recursion}
\tau_{\dpol}(s,\bar{s}) &= t_{\dpol}(s)  + \sum_{s'\in{\mathcal{B}}_s}p_{\dpol}(s,s')\tau_{\dpol}(s',\bar{s})\,,
\end{align}
The relation Eq.(\ref{eq:recursion}) 
is supplemented with the trivial boundary condition $\tau_\dpol(\bar{s},\bar{s})=0$.

%%%%%%%%%%%%%%%%%%%%%%%%%%%%%%%%%%%%%%%%%%%%%%%%%%
\section{Dynamic Programming}
%%%%%%%%%%%%%%%%%%%%%%%%%%%%%%%%%%%%%%%%%%%%%%%%%%
\label{s:DP}

Our goal now is to determine the optimal policy 
\begin{align}
\dpol_*= \mathrm{argmin}_{\dpol}\tau_{\dpol}(s,\bar{s})
\end{align}
that minimizes the time $\tau_\dpol(s,\bar{s})$ to reach the target $\bar{s}$
from each and every state $s$.
The resulting optimal first passage time is 
\begin{align}
\tau_*(s,\bar{s}) = \min_{\dpol}\tau_{\dpol}(s,\bar{s}).
\end{align}
Such a optimization problem is called a Markov decision process (MDP).
Using well-known methods of MDPs, 
we substitute the optimal policy in Eq.(\ref{eq:recursion}) to obtain
the so-called Bellmann optimality equation~\cite{Bellman1957,Sutton1998}
\begin{align}\label{eq:Bellman}
\tau_*(s,\bar{s}) = \min_{\dpol(s)}\Bigl[  
t_{\dpol}(s)  + \sum_{s'\in {\mathcal{B}}_s}p_{\dpol}(s,s')\tau_*(s',\bar{s})
\Bigr] \,.
\end{align}
This equation can be solved numerically using a dynamic programming method called value iteration~\cite{Sutton1998},
which consists in substituting an estimate of  $\tau_*(s')$ in the right hand side of Eq.(\ref{eq:Bellman})
to obtain an improved estimate of $\tau_*(s)$ at each iteration.

This method requires to list all states in the system. 
As a consequence, it is suitable for small clusters.
Indeed, the number $S_N$ of cluster configurations grows exponentially
with the number $N$ of particles.
These configurations are sometimes called free polyominoes or lattice animals
in the literature.
The asymptotic behavior of $S_N$ for large $N$ is 
\begin{align}
    S_N \sim c\lambda^N/N,
    \label{eq:SN_asymptotic}
\end{align}  
with $ \lambda \approx 4.0626$ and $ c \approx 0.3169$~\cite{Jensen2000}.
Due to memory limitation, our simulations
are performed with $N\leq12$. Hence, the biggest
clusters that we have studied have $S_{12}\approx 5\times 10^{5}$ states.
We have used a Python implementation of the algorithm presented in Ref.~\cite{haskell_polyominoes} 
to generate all possible  polyominoes for a given value of $N$~\footnote{
More specifically, the polyominoes that are suitable for our analysis
are the so-called {\it free} polyominoes.}.
Note also the double-exponential increase of the number of policies $3^{S_N}$
which reaches quickly very large numbers, e.g. $\sim 10^{10^5}$ for $N=12$,
forbidding any direct solution of the minimization problem
based on exhaustive listing and evaluation
of all policies.

Although we compute the optimal times $\tau_*(s,\bar{s})$ which 
are different for each state $s$,
we choose to focus the analysis on a single quantity that depends 
only on the target state: 
the expected return time to target $\tau_\dpol^\mathrm{r}(\bar{s})$.
This quantity is defined as the expected time spent outside the target 
for dynamical trajectories that start from the target and return for the first time
to the target~\cite{Boccardo2022a}
\begin{align}
\label{eq:mean_return_time}
\tau_\dpol^\mathrm{r}(\bar{s}) &= \sum_{s\in{\mathcal{B}}_{\bar{s}}}p_{\dpol}(\bar{s},s)\tau_\dpol(s,\bar{s})
\, .
\end{align}
Note that such a definition requires to extend the definition of the policy.
Indeed, we need to define the value of the force on the target state itself.
However, due to the Markovian character of the dynamics, 
this additional force on the target does not affect the mean first passage time to target
from the other states outside the target (and as a consequence, it does
not affect the optimal policy).
We choose to set the force on the target so as to minimize the expected return time $\tau_\dpol^\mathrm{r}(\bar{s})$.
On a computational level,
we define a new state $\hat{s}$ which is an artificial copy 
of the target state $ \bar{s} $, with the same configuration than the target but 
with a residence time equal to zero.
The Bellman equation for the state $\hat{s}$ therefore reads
\begin{align}
\tau_*(\hat{s},\bar{s}) = \min_{\dpol(s)}\Bigl[  
\sum_{s'\in {\mathcal{B}}_{\bar{s}}}p_{\dpol}(s,s')\tau_{\dpol}(s',\bar{s})
\Bigr] \,.
\end{align}
We then have $\tau_*^\mathrm{r}(\bar{s})=\tau_*(\hat{s},\bar{s})$.

%%%%%%%%%%%%%%%%%%%%%%%%%%%%%%%%%%%%%%%%%%%%
%%%%%%%%%%%%%%%%%%%%%%%%%%%%%%%%%%%%%%%%%%%%
\section{Temperature dependence of the return time to target }
%%%%%%%%%%%%%%%%%%%%%%%%%%%%%%%%%%%%%%%%%%%%
%%%%%%%%%%%%%%%%%%%%%%%%%%%%%%%%%%%%%%%%%%%%

The temperature dependence of the 
expected return time to target $\tau_\dpol^\mathrm{r}(\bar{s})$ 
was discussed in our previous paper with or without force~\cite{Boccardo2022a,Boccardo2022b}.
In this section, we briefly report the main features of this temperature-dependence.

Furthermore, although the limit of very high temperatures $k_BT/J\gg1$ is not described
accurately by the Arrhenius form of the hopping rate \cref{eq:hopping_rate} as discussed above,
the formal study of this limit provides important information on
the behavior at finite temperatures.
Indeed, the energies are irrelevant when $k_BT/J\gg1$, 
and an exact expression can be derived ${\tau}^\mathrm{r}_\infty(\bar{s})=(S_N-1)/d_{\bar{s}}$,
where the degree $d_{\bar{s}}$ of the target $\bar{s}$ 
is the number of states that can be reached from the target in one move
(i.e., $d_{\bar{s}}$ is the cardinal of ${\cal B}_{\bar s}$).

This high temperature limit is independent of the 
external force. Hence, there is no possible gain in the time
to reach a target in the high temperature limit.
However, we showed that the optimization of the forces allows one to 
decrease the time to reach the target at finite temperatures. This optimization gain increases
when the temperature decreases~\cite{Boccardo2022a}. 

Furthermore, from the high-temperature limit,  $\tau_*(s,\bar{s})$
and $\tau_*^\mathrm{r}(\bar{s})$
can increase or decrease as the temperature is decreased.
A decrease of these quantities is obtained when the cluster is large enough and when the target
is a compact shape.
In contrast, at low temperatures, $\tau_*(s,\bar{s})$
and $\tau_*^\mathrm{r}(\bar{s})$ always increase when the temperature is 
decreased because the rates for particle hopping become very small.

In the present paper, we do not focus on the 
dependence of $\tau_*(s,\bar{s})$ and $\tau_*^\mathrm{r}(\bar{s})$
on temperature. Instead, we analyze the properties
of the optimal policy $\dpol_*$ itself.

%%%%%%%%%%%%%%%%%%%%%%%%%%%%%%%%%%%%%%%%%%%%
%%%%%%%%%%%%%%%%%%%%%%%%%%%%%%%%%%%%%%%%%%%%
\section{Small clusters}
%%%%%%%%%%%%%%%%%%%%%%%%%%%%%%%%%%%%%%%%%%%%
%%%%%%%%%%%%%%%%%%%%%%%%%%%%%%%%%%%%%%%%%%%%

%%%%%%%%%%%%%%%%%%%%%%%%%%%%%%%%%%%%%%%%%%%%
%%%%%%%%%%%%%%%%%%%%%%%%%%%%%%%%%%%%%%%%%%%%
\subsection{Dimers}
%%%%%%%%%%%%%%%%%%%%%%%%%%%%%%%%%%%%%%%%%%%%
%%%%%%%%%%%%%%%%%%%%%%%%%%%%%%%%%%%%%%%%%%%%

\begin{figure}[ht]
    \includegraphics[width=.5\linewidth]{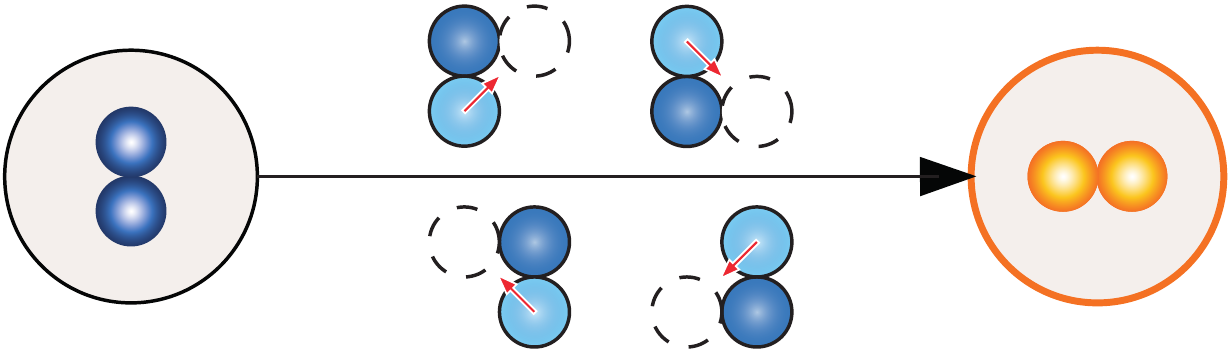}
    \caption{Graph of configurations of a dimer, with the only 4 possible particle moves to reach the orange target. Top two moves have $ u_{ss'}=1/2 $ while the two bottom ones have $ u_{ss'}=-1/2 $.}
    \label{fig:graph_dimer}
\end{figure}

The simplest case is that of a dimer cluster.
The graph of configurations of a dimer only consist of $S_2=2$
states. The first one is along $\hat{\bm y}$, and the
second one is along $\hat{\bm x}$.
Let us consider the case where the horizontal dimer along $\hat{\bm x}$ is the target $\bar{s}$,
as in \cref{fig:graph_dimer}.
This situation is simple enough to allow for an
explicit derivation of the optimal policy.

Since  $ \tau_{\dpol}(\bar{s},\bar{s})=0 $ on the target,
the recursion equation (\ref{eq:recursion}) shows that
the expected time to reach the target is equal 
to the residence time $ \tau_{\dpol}(s, \bar{s}) = t_{\dpol}(s) $. 
This quantity is also equal to the expected return time to the target ${\tau}^\mathrm{r}_\infty(\bar{s})$
from Eq.(\ref{eq:mean_return_time}).
There are  4 possible  moves that take the system  from $ s $ to $ \bar{s} $.
 This is actually a specific property of dimers. Indeed, for larger clusters with $N>2$,
there is always a unique move to go from one state to another state~\cite{Boccardo2022a,Boccardo2022b}.
The moves are shown in \cref{fig:graph_dimer}. 
Two moves are in the $+\bm{x}$ direction and have $ u_{ss'}=1/2 $ 
while the other two are in the $-\bm{x}$ direction and have $ u_{ss'}=-1/2 $.
Hence, from Eq.(\ref{eq:mean_res_time}) we have
\begin{align}
	{\tau}^\mathrm{r}_\infty(\bar{s})=\tau_{\dpol}(s, \bar{s}) = t_{\dpol}(s)
	=\frac{{\rm e}^{1/T}}{4\cosh(\varphi(s)/2T)}
	\label{eq:dimer_tau}
\end{align}
where 
$\varphi(s)$ is the scalar force in state $s$, so that $\dpol(s)=\varphi(s)\,\hat {\bm x}$.

As a first remark, $\tau_{\dpol}(s, \bar{s})$ is always increasing when decreasing
temperature (as long as $F_0<2$ for the energy barriers to be positive
as discussed in Sec.~\ref{s:model}). Such a behavior is expected
for small clusters as discussed in Ref.~\cite{Boccardo2022a,Boccardo2022b}.

Moreover by symmetry, and as can be seen from Eq.(\ref{eq:dimer_tau}), 
both choices $\varphi(s)=\pm F_0$  result in the same expected time to target 
$ \tau_{\dpol}(s, \bar{s}) $. These choices are both optimal, 
because the other choice of a vanishing force $\varphi(s)=0$ leads to a larger
value of $\tau_{\dpol}(s, \bar{s})$. 
Thus, for the dimer, we have two equivalent optimal policies $ \dpol_*(s)=+ F_0 \hat{\bm{x}} $,
or $ \dpol_*(s)=- F_0 \hat{\bm{x}} $.

As a summary of the trivial dimer case, we find that 
there are two equivalent optimal policies, with a non-zero-force 
in the $\pm \hat{\bm{x}}$ direction. We call degeneracy the fact of having a 
non-unique optimal policy.
In the dimer case, degeneracy is a consequence
of the concomitant invariance of the two cluster shapes
and of the set of possible forces
under the symmetry transformation $x \rightarrow -x$.

%%%%%%%%%%%%%%%%%%%%%%%%%%%%%%%%%%%%%%%%%%%%
%%%%%%%%%%%%%%%%%%%%%%%%%%%%%%%%%%%%%%%%%%%%
\subsection{Trimers}
%%%%%%%%%%%%%%%%%%%%%%%%%%%%%%%%%%%%%%%%%%%%
%%%%%%%%%%%%%%%%%%%%%%%%%%%%%%%%%%%%%%%%%%%%
\label{s:trimers}

\begin{figure}[ht]
    \includegraphics[width=.5\linewidth]{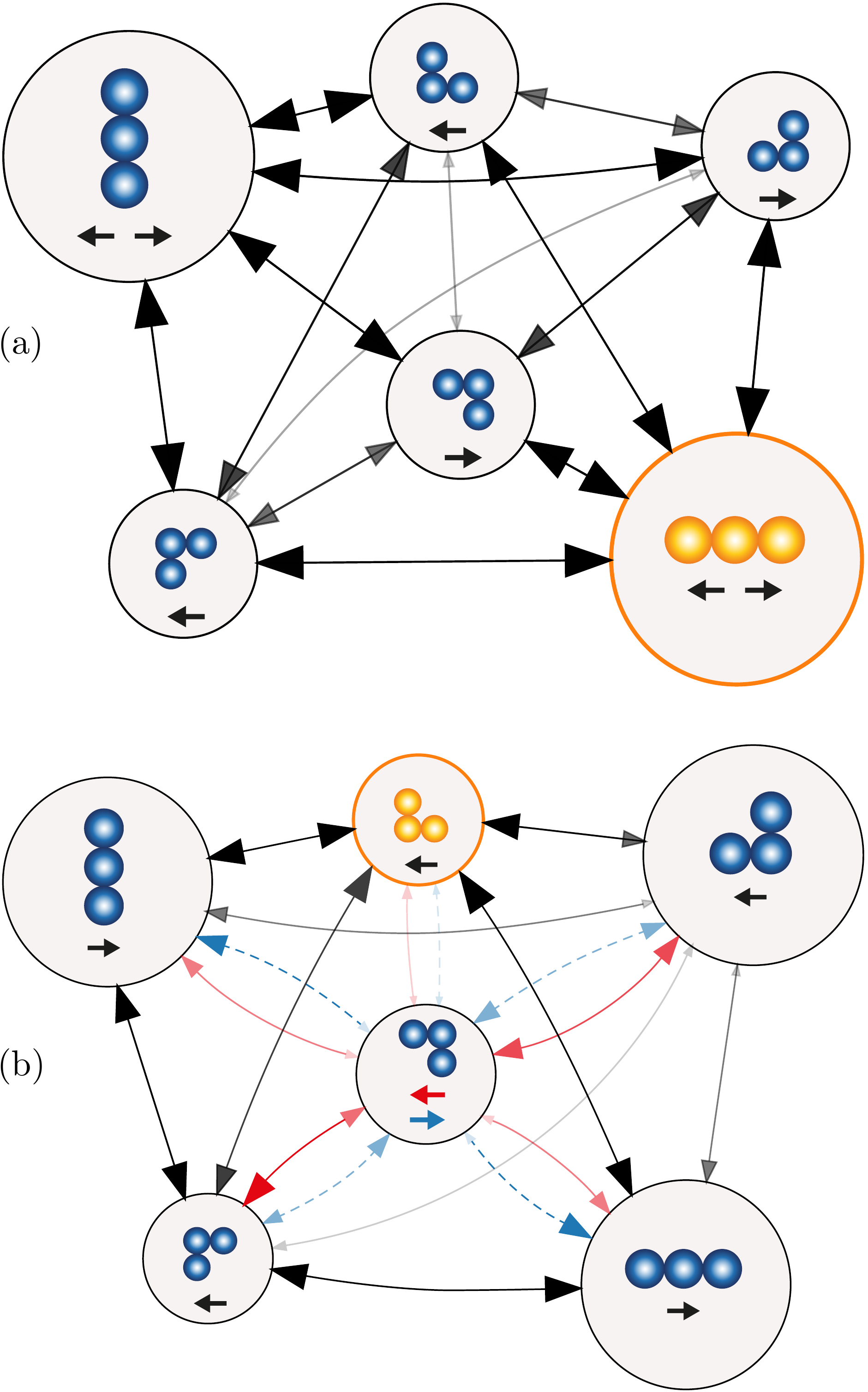}
    \caption{Two graphs of the trimer dynamics (N=3), with the optimal policy computed for two different targets (in orange). (a) $T=0.61$, in the upper left state, the optimal action is degenerate, and the force can be equivalently set to the right or left. (b) Two optimal policies at different temperatures: $T=0.66$ (in blue) and $T=0.67$ (in red).  The optimal policy changes only
    in one state (in the center) when the temperature is increased.
	}
	\label{fig:graph_2_11}
\end{figure}

After the dimer, the simplest 
clusters are trimers, i.e., clusters with $N=3$ particles and $S_3=6$ states.
Despite their small size, a direct analytical solution is cumbersome, and we resort to the numerical methods
described in \cref{s:DP}.
A schematic of the optimal policy is shown in \cref{fig:graph_2_11} for two different targets.

A first observation is that degeneracy can also be
found in the optimal policy for trimers, as seen in \cref{fig:graph_2_11}(a)
where the target state is the configuration  aligned along the $x$ axis.
While the states that are not compatible with the 
$x\rightarrow -x$ symmetry exhibit a unique optimal choice
of the external force, the state with  the three particles aligned
along the $y$ axis presents a degenerated force.
Again, the degeneracy is a clear consequence of the 
compatibility between the symmetries of the configurations and
the symmetries of the force.
However, a well known property of the solution $\tau_*(s,\bar{s})$
of the Bellmann optimality equation~\cref{eq:Bellman} is its unicity~\cite{Sutton1998}.
Hence, the two degenerate policies in \cref{fig:graph_2_11}(a)
lead to the same value of the the first passage times $\tau_*(s,\bar{s})$.

Furthermore, as opposed to the case of dimers,
the optimal policy  for trimers  $ \dpol_* $ can change when the temperature is varied. 
An example of policy transition
for the trimer is shown in \cref{fig:graph_2_11}(b). 
In the state at the center of the graph, the optimal action flips from right to left at $T=T_c$, with $0.66<T_c<0.67$.
The optimal policies above and below the transition at $T\leq 0.66$ and $T\geq 0.67$
are represented in \cref{fig:graph_2_11}(b).
The difference between the high-temperature and the low-temperature optimal
policies consist in the change of orientation of the force in a single state.

\begin{figure}[t!]
    \includegraphics[width=.5\linewidth]{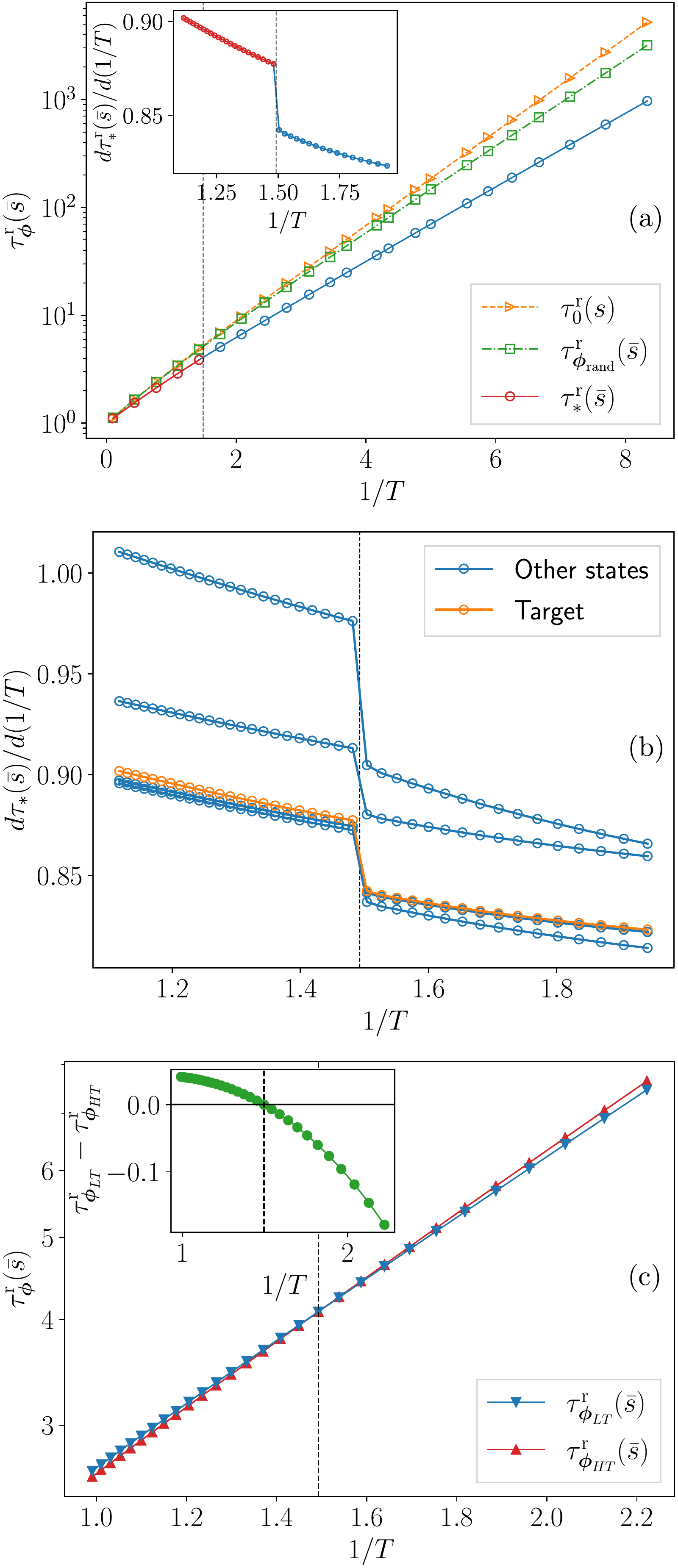}
    \caption{(a) Return time to target 
    as a function of $1/T$ 
    for the optimal policy $\dpol_*$ for the trimer target of \cref{fig:graph_2_11}(b).  
    For comparison, we also report the return time for the random policy $\dpol_{\textrm{rand}}$ and the zero-force policy $\dpol(s)=0$
    (defined and evaluated in \cite{Boccardo2022a}). Inset: first derivative of the optimal return time near the transition temperature $T_c$ (vertical line). (b) Derivative of the expected optimal first passage time to target starting from all other states in the system. (c) Expected return time to target for the high- and low-temperature optimal policies. Inset: their difference.
    }
\label{fig:plots_trimer}
\end{figure}

The mean return time to target $\tau_*^\mathrm{r}(\bar{s})$ for the trimer target of \cref{fig:graph_2_11}(b)
is reported in \cref{fig:plots_trimer}(a). The transition at $T_c$ is seen to be continuous 
for $\tau_*^\mathrm{r}(\bar{s})$.
The continuity of $\tau_*^\mathrm{r}(\bar{s})$ means that the low-temperature and the high-temperature
policies have the same performance at the transition. However, 
the first derivative with respect to the temperature are discontinuous,
as shown in the inset of \cref{fig:plots_trimer}(a).
Indeed, the derivatives
are properties of the policies themselves, and are therefore expected to be 
different on the left and right sides of the transition.
The first passage times $\tau_*(s,\bar{s})$ to the target 
from the other 5 states outside the target exhibit the
same discontinuity and their derivative are shown in \cref{fig:plots_trimer}(b).

As an additional remark, the discontinuity of the derivative 
is always negative. Indeed, let us denote the high-temperature
and low-temperature optimal policies on both sides
of the transition as $\dpol_{HT}$ and $\dpol_{LT}$.
The discontinuity corresponds to a 
crossing of the functions ${\tau}^{\mathrm{r}}_{\dpol_{HT}}(\bar{s})$
and ${\tau}^{\mathrm{r}}_{\dpol_{LT}}(\bar{s})$ that must be optimal
at high and low temperatures respectively. As seen in \cref{fig:plots_trimer}(c), the 
policy $\dpol_{LT}$ on the low-temperature side must have a smaller slope than $\dpol_{HT}$ as a 
function of the inverse temperature. As a consequence,
the jump of the derivative of  $\tau_*^\mathrm{r}(\bar{s})$
with respect to $1/T$ is always negative.

%%%%%%%%%%%%%%%%%%%%%%%%%%%%%%%%%%%%%%%%%%%%%%%%%%%
%%%%%%%%%%%%%%%%%%%%%%%%%%%%%%%%%%%%%%%%%%%%%%%%%%%
\subsection{Tetramers}
%%%%%%%%%%%%%%%%%%%%%%%%%%%%%%%%%%%%%%%%%%%%%%%%%%%
%%%%%%%%%%%%%%%%%%%%%%%%%%%%%%%%%%%%%%%%%%%%%%%%%%%
\label{s:tetramers}

\begin{figure}[ht]
	\includegraphics[width=.5\linewidth]{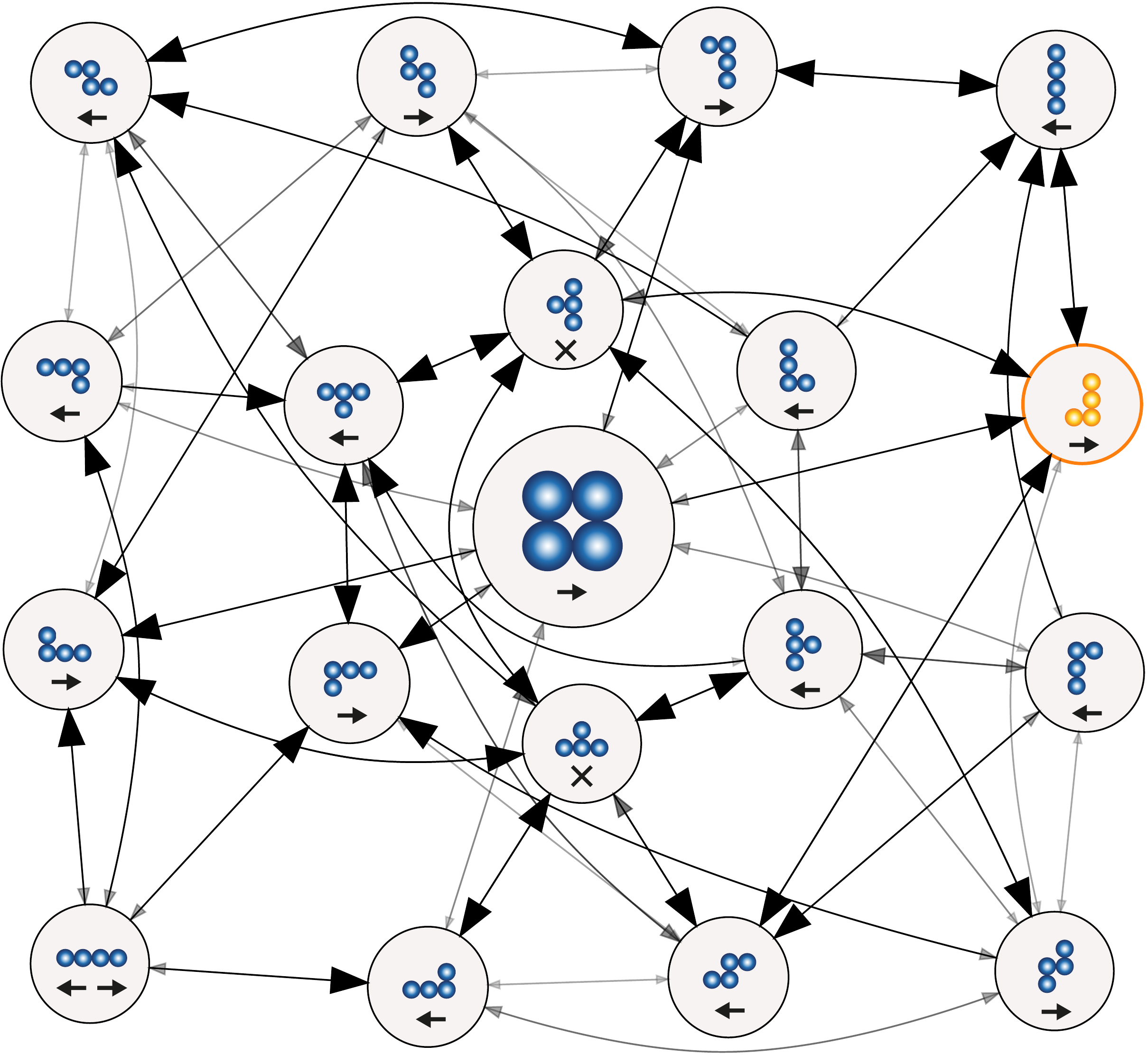}
	\caption{Dynamical graph of a tetramer cluster at $T=0.24$, with the optimal policies to reach the orange target. This target belongs to the $\CMcal{N}$ symmetry class, however, the state with the 4 particles aligned along the $x$ direction (bottom left) shows a degenerate optimal action.
	}
	\label{fig:graph_tetramer_nontrivial}
\end{figure}

The case of tetramers with $N=4$ and $S_4=19$ states
exhibits two novelties. First, degeneracy is observed
in cases where the target does not exhibit the $x\rightarrow -x$ symmetry of the force.
One example of this situation
is shown in \cref{fig:graph_tetramer_nontrivial}. 
The state with the 4 particles aligned along the $x$ direction is the only degenerate state. 
Two other examples of tetramer targets that do not have the $x\rightarrow -x$ symmetry, 
but that exhibit degenerate states are reported in \cref{fig:graph_tetramer_nontrivial_app} 
of the Appendix. 
We interpret the presence of such degenerate states
in the absence of obvious symmetries of the target
as a consequence from 
the existence of some non-trivial symmetry of the full dynamical graph.
However, we have not identified this symmetry explictely.

The second difference is that several temperature
transitions are possible for tetramers, while we have observed at most one for
trimers and none for dimers.
In the temperature range that we have considered $0.1<T<0.6$, 
there are between $1$ and $10$ transitions for tetramers
depending on the target.

%%%%%%%%%%%%%%%%%%%%%%%%%%%%%%%%%%%%%%%%%%%%%%%%%%%
%%%%%%%%%%%%%%%%%%%%%%%%%%%%%%%%%%%%%%%%%%%%%%%%%%%
\section{Larger clusters}
%%%%%%%%%%%%%%%%%%%%%%%%%%%%%%%%%%%%%%%%%%%%%%%%%%%
%%%%%%%%%%%%%%%%%%%%%%%%%%%%%%%%%%%%%%%%%%%%%%%%%%%

The above results for dimers, trimers and tetramers raise
questions on the behavior of the optimal policy for larger clusters.
A first question is the number of degenerate states and their relations to
cluster symmetries. A second one is the increase of the 
number of transitions with cluster size. These two questions are discussed below.

%%%%%%%%%%%%%%%%%%%%%%%%%%%%%%%%%%%%%%%%%%%%%%%%%%%
\subsection{Degeneracy}
%%%%%%%%%%%%%%%%%%%%%%%%%%%%%%%%%%%%%%%%%%%%%%%%%%%

Let us consider an optimal policy for the target $\bar{s}$, 
which associates a force $\dpol_*(s)$ to each state $s$.
Performing one of the transformations 
${x}\rightarrow{-x}$, ${y}\rightarrow{-y}$, or $({x}, {y}) \rightarrow({-x},{-y})$
on all the states and forces, we obtain another optimal policy.
If the target $\bar{s}$ is transformed in itself,
then the new policy is again an optimal policy for $\bar{s}$.
If in addition a state $s$ is transformed in itself
and the force in $s$ is transformed into its opposite,
then the force $\dpol_*(s)$ and its opposite $-\dpol_*(s)$ are both optimal,
and there are two possibilities. First possibility, the force
in the state $s$ vanishes. Indeed, it is clear
that $\dpol_*(s)=0$ is equal to its opposite.
Second possibility, the force is non-zero. Then, 
both $\dpol_*(s)=F_0\hat{\bm x}$ and $\dpol_*(s)=-F_0\hat{\bm x}$
are solutions, and the state is called a degenerate state.

Note that the above statements correspond
to switching all the forces in the graph simultaneously.
Hence, this does not exclude the possibility of 
correlations of the forces in different degenerate states
(for example, if an optimal policy sets the forces in the same $+\hat{\bm{x}}$ direction
in two degenerate states, then we have shown that there
is another optimal policy with the two forces in the $-\hat{\bm{x}}$ direction,
but we have not shown that the policy with the two forces in 
opposite directions is also optimal). 
However, we can make use of a well known property of the optimal solution
of \cref{eq:Bellman} already mentioned in \cref{s:trimers}: in contrast to the optimal policy $\dpol_*(s)$ which is
not unique, the value of $\tau_*(s,\bar{s})$ is unique~\cite{Sutton1998}.
Hence, switching the force in degenerate states does not change $\tau_*(s,\bar{s})$. 
In addition, the quantities 
$t_\dpol(s)$ and $p_\dpol(s,\bar{s})$ that enter into \cref{eq:Bellman} depend
only of the force $\dpol(s)$ in the state $s$. Hence,
\cref{eq:Bellman} involves only the force in $s$, and
the optimal policy $\dpol_*(s)$ 
cannot depend on the optimal force $\dpol_*(s')$ in other states $s'\neq s$.

As a summary, if a target $\bar{s}$ and a state $s$
are both invariant under a symmetry transformation
and if the force in $s$ is reversed under this transformation,
then the optimal policy in state $s$ has either a vanishing force,
or a degenerate force which can be switched independently
from the other forces in other degenerate states.
The property that in a given
degenerate state, the force can be switched independently
from the other degenerate states was confirmed numerically.

\begin{figure}[ht]
% 	\centering
%	\includegraphics[width=\linewidth]{figure7}
	\includegraphics[width=.5\linewidth]{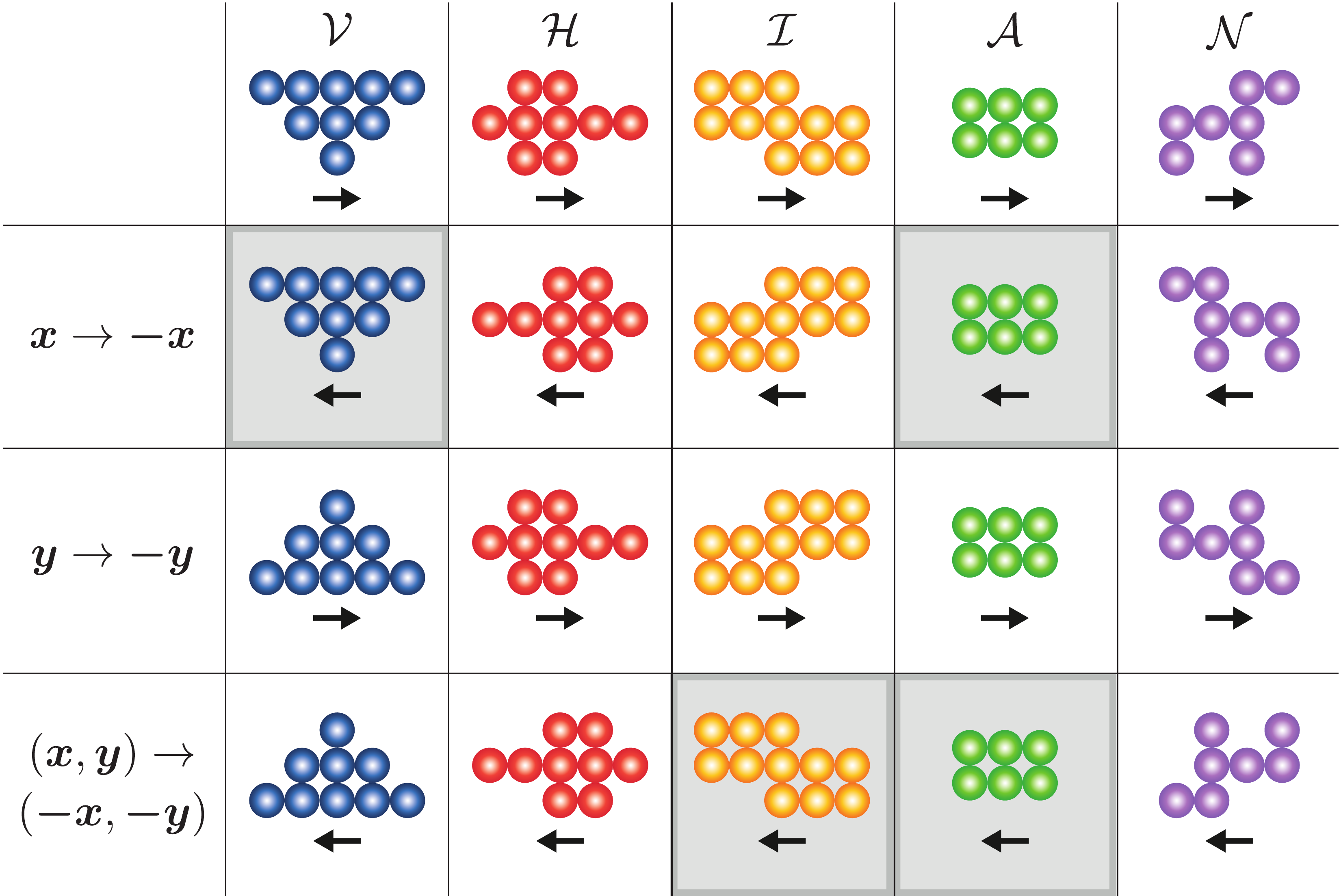}
	\caption{The five symmetry classes and their transformations.}
	\label{fig:symm_examples}
\end{figure}

To investigate the relation between degeneracy of the optimal actions
and symmetries in more details,
we define five mutually exclusive symmetry classes of states.
One example of each class is shown in the first line of \cref{fig:symm_examples}.
The first class $\CMcal{V}$ includes clusters with 
    mirror symmetry with respect to a Vertical ${y}$ axis,
    or invariance under the ${x}\rightarrow{-x}$ transformation.
The second class $\CMcal{H}$ includes clusters with 
mirror symmetry with respect to a Horizontal ${x}$ axis,
    or invariance under the ${y}\rightarrow{-y}$ transformation.
The third class $\CMcal{I}$ corresponds to clusters with
space Inversion symmetry,
    or invariance under the transformation $({x}, {y}) \rightarrow({-x},{-y})$.
The fourth class $\CMcal{A}$ includes clusters 
    that are separately invariant  under
    the ${x}\rightarrow{-x}$ transformation 
    and under the ${y}\rightarrow{-y}$ transformation.
The fifth class $\CMcal{N}$ corresponds to clusters with None of the above symmetries.
These five classes cover all cluster configurations and have no overlap, 
in the sense that each shape belongs to only one class, 
corresponding to the class with the highest symmetry where it can be included. 

Let us consider how states belonging to these five symmetry classes 
and the associated optimal force
are transformed when the entire dynamical graph of the system and the force 
are transformed under one of the three transformations 
${x}\rightarrow{-x}$, ${y}\rightarrow{-y}$, or $({x}, {y}) \rightarrow({-x},{-y})$. 
This is illustrated in the table of \cref{fig:symm_examples}. 
Among the 15 cases in the table, 
four cases have a transformed shape which is identical
to the initial one, with a force that is flipped.
These 4 cases, which correspond either to 
a vanishing or to a degenerate optimal force, 
have a background colored in grey in the table.

More precisely, we expect the following symmetry rules. For a 
$\CMcal{V}$
target, all states that are invariant with respect to the ${x}\rightarrow{-x}$ transformation, 
i.e. the 
$\CMcal{V}$ and $\CMcal{A}$
states, will be either degenerate or with a zero-force optimal action. Instead, for a 
$\CMcal{I}$
target, all states that are invariant with respect to the $({x}, {y}) \rightarrow({-x},{-y})$ transformation, i.e. the
$\CMcal{I}$ and $\CMcal{A}$
states, will be either degenerate or with a zero-force optimal action. Finally, for a target in the
$\CMcal{A}$
symmetry class, all states that are invariant with respect to either the 
${x}\rightarrow{-x}$, 
or the $({x}, {y}) \rightarrow({-x},{-y})$ transformation, i.e. the
$\CMcal{V}$, $\CMcal{I}$ and $\CMcal{A}$
states, will be either degenerate or with a zero-force optimal action.

We have checked the validity of the above symmetry rules
for all possible targets with $N\leq7$ ($ S_7=760 $):
states belonging the above mentioned classes indeed have either 
a vanishing or a degenerate optimal force. 
Due to computational limitations, we were able to check only some 
arbitrarily selected targets 
with $ 8\leq N\leq 12 $. 
Again, all states obeying
the symmetry rules mentioned above exhibit a vanishing or degenerate 
optimal force.

However, when a state do not obey these  symmetry rules, we
have no indication about its possible degeneracy.
Indeed, other \q{hidden} symmetries of the graph could come into play.
We found only
10 cases of degenerate states outside these symmetry rules,
and they all correspond to the tetramer ($N=4$) cases discussed in \cref{s:tetramers}. 
Since, we have checked all targets 
with $N\leq 7$, we find that they are the only cases for $N\leq 7$.
However we do not know if there are other cases for $ 8 \leq N \leq 12 $, 
since we could not check all possible targets in this range.

\begin{figure}[ht]
    \includegraphics[width=.5\linewidth]{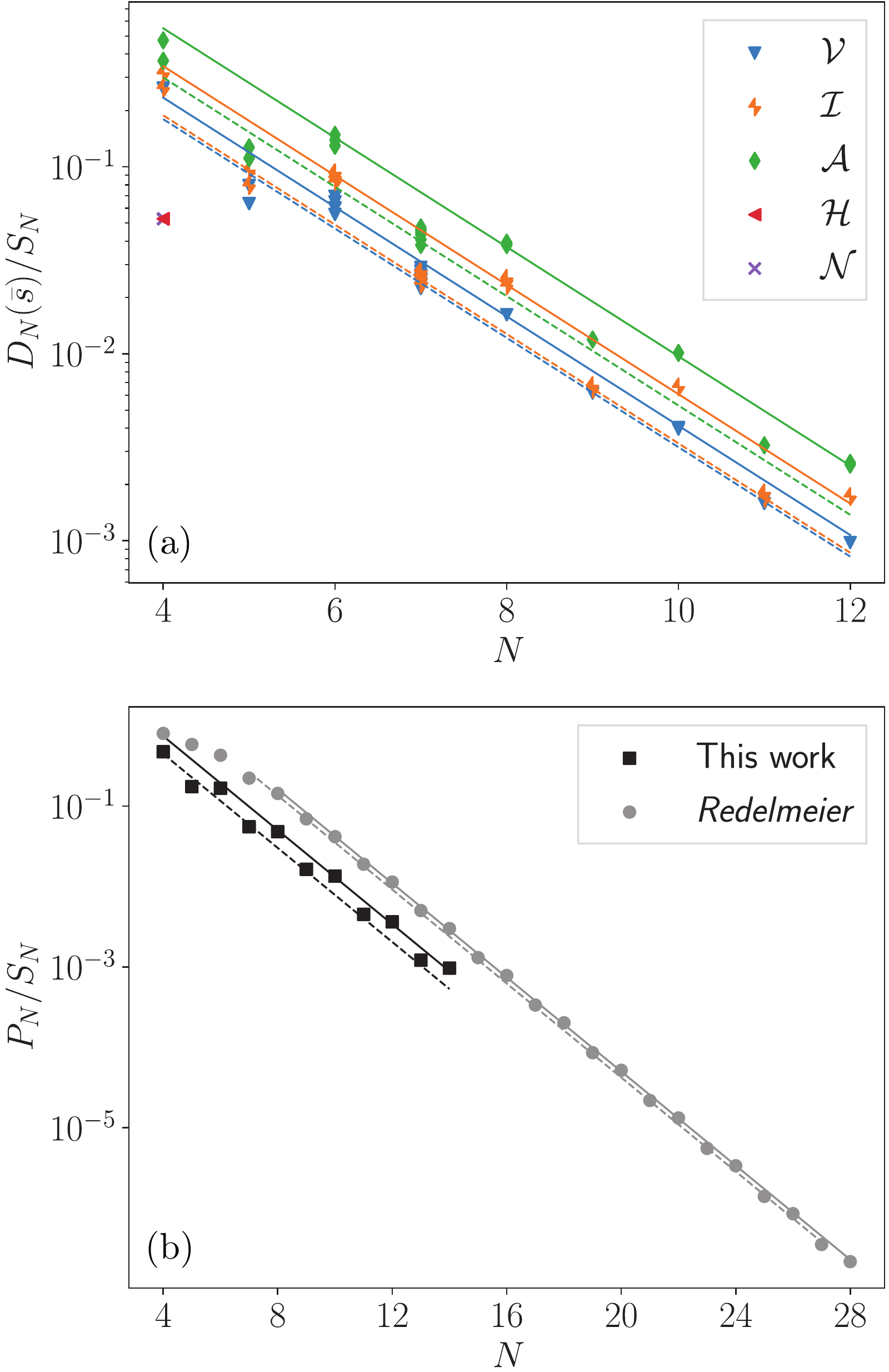}
    \caption{(a) Fraction of degenerate states for several targets at a fixed temperature $ T = 0.24 $. 
    Different symbols (colors) correspond to targets belonging to different symmetry classes.
    (b) Total fraction of symmetric polyominoes as a function of $ N $. 
    Our results consider only $\CMcal{V}$, $\CMcal{I}$ and $\CMcal{A}$ symmetry classes, while Redelmeier's results include more classes.
    The lines in (a) and (b) correspond to the scaling $\sim \mu^N$, with $\mu \approx 0.51$ obtained fitting Redelmeier's data. 
}
    \label{fig:degeneracy_plots}
\end{figure}

To investigate the frequency of appearance of degenerate states, 
we define $ D_N(\bar{s}) $, the number of degenerate states in 
the optimal policy of a target $ \bar{s} $ of size $N$. 
The fraction of degenerate states $ D_N(\bar{s})/S_N $ for 
various targets of size $ 4 \leq N \leq 12 $ 
is plotted in \cref{fig:degeneracy_plots}(a) for a fixed
temperature $T=0.24$. 
In \cref{fig:degeneracy_plots}(a), different symbols (colors)
represent the symmetry classes of the target.
We have reported all targets with $ 4 \leq N \leq 7 $,
and some arbitrary selected targets with $ 8 \leq N \leq 12 $.
Beyond the exceptions at $ N=4 $, 
the fraction $ D_N(\bar{s})/S_N $ vanishes for 
the $\CMcal{H}$ and $\CMcal{N}$ classes.
In contrast, the targets belonging to the other classes of symmetry
exhibit a non-zero fraction of degenerate state. 
 We also observe that the variation
of the values of the fraction of degenerate states
for different targets with the same $N$ is  small 
as compared to the variation with $N$. 
We therefore conclude that the most relevant parameter
for the variation of the fraction of degenerate states
is $N$.

The fraction $D_N(\bar{s})/S_N$ for targets that belong
to $\CMcal{V}$, $\CMcal{I}$, or $\CMcal{A}$ symmetry classes is found to decay exponentially with $N$.
Let us define $ P_N $, the number of polyominoes of
size $ N $ belonging to one of the three symmetry classes 
$\CMcal{V}$, $\CMcal{I}$ or $\CMcal{A}$. 
In \cref{fig:degeneracy_plots}(b), 
the fraction of symmetric polyominoes $ P_N/S_N $ (for $ 4\leq N \leq 14 $) is found to exhibit the same exponential decay as Redelmeier's results from Ref.~\cite{Redelmeier1981} up to $ N=24 $ (and extended to $ N=28 $ by Oliveira e Silva~\cite{OliveiraeSilva2015}). 
Note that the fractions of symmetric polyominoes computed by
Redelmeier are larger than ours because they take into account
more symmetry classes, however the scaling of this fraction
with respect to $N$
is the same $\sim \mu^N$, where  $\mu\approx 0.51$
(the fit was done by considering odd and even values of $ N $ separately). The fraction of degenerate states 
for a given symmetry class reported in \cref{fig:degeneracy_plots}(a) 
exhibits the same behavior $D_N(\bar{s})/S_N\sim \mu^N$.
A value $\mu<1$ indicates that $D_N(\bar{s})/S_N\rightarrow 0$ for large $N$. However, the number of degenerate states in a given class $D_N(\bar{s})\sim (\lambda \mu)^N$ with $\lambda \mu \approx 2.07>1$ grows exponentially with $N$.

As stated above, states that obey
the symmetry rules are also compatible with a zero force
in the optimal policy. 
In \cref{fig:graph_tetramer_symmetric} in Appendix, we report a dynamical graph where this is the case.
Thus, the number of degenerate states is not fixed by symmetry and can vary with temperature. 
However, we have checked that $D_N(\bar{s})/S_N$ only varies weakly with temperature. 
Hence, the results discussed above on the variation of $D_N(\bar{s})/S_N$ with $N$ at $T=0.24$ 
are expected to hold at other temperatures.

As a summary, two conclusions can be drawn from this analysis
of degenerate states. First, most degeneracies in the optimal 
actions are associated to symmetries of the target state, 
and not to nontrivial symmetries of the graph representing the full dynamical system. 
Second, although the number of degenerate states increases exponentially, 
the fraction of degenerate states vanishes as $N$ increases.

%%%%%%%%%%%%%%%%%%%%%%%%%%%%%%%%%%%%%%%%%%%%%%%%%%%
\subsection{Transition density}
%%%%%%%%%%%%%%%%%%%%%%%%%%%%%%%%%%%%%%%%%%%%%%%%%%%

\begin{figure}[h!t]
	\includegraphics[width=.5\linewidth]{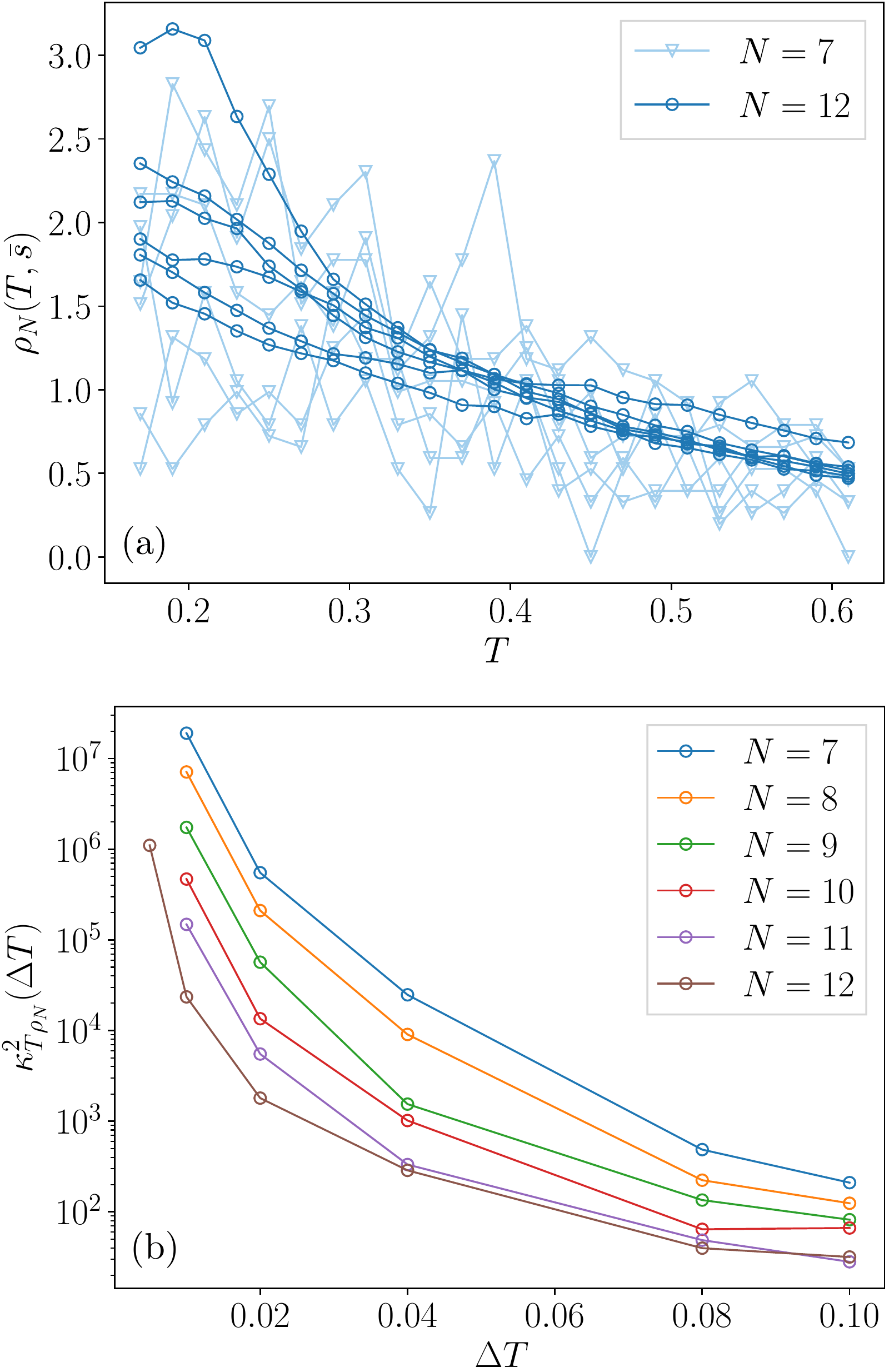}
	\caption{(a) Density of change $ \rho_N (T, \bar{s}) $ as a function of $T$ for several targets of size 7 and 12 (each curve corresponds to a different target). The step in the temperature is $\Delta T=0.02$. (b) Local mean square curvature $\kappa^2_{T\rho_N}(\Delta T)$ as a function of $\Delta T$ for different values of $N$, averaged over several targets. Averages are performed over 10 targets for $N=7$, to 4 targets for $N = 10$. For $\Delta T=0.005$ and $N=12$, the average is performed on 2 targets and in a smaller temperature range $0.21 \leq T \leq 0.3$.}
	\label{fig:density}
\end{figure}

We now turn to the analysis of the growth of the 
number of temperature transitions in the optimal policy
when increasing the cluster size $N$.
This increase is strong and as $N$ reaches 12, the number
of transitions is so large that identifying each single 
transition is clearly not a meaningful approach.
We therefore resort to a statistical analysis of the transitions.
We define the density of change 
\begin{equation}
    \rho_N (T, \bar{s})=\frac{ \Delta S_N (T, \bar{s})}{S_N \Delta T }, 
\end{equation}
where $\Delta S_N (T, \bar{s})$ is the number of states where the optimal action has changed between $ T $ and $T + \Delta T $, for a target $ \bar{s} $ of size $N$. In \cref{fig:density}, we show $\rho_N(T, \bar{s})$ with $ \Delta T = 0.02$ for several targets of size 7 and 12.

The first important feature that emerges from this plot is that 
$\rho_N (T, \bar{s})$ is of the order of one. 
Therefore using \cref{eq:SN_asymptotic},
we  can conclude that $\Delta S_N (T, \bar{s})/\Delta T$ 
grows exponentially like $S_N$ when $N$ increases. 

Second, $\rho_N (T, \bar{s})$ becomes smoother for $N=12$, 
suggesting that the density of transitions tends to a well defined 
smooth function of $T$ as $N$ increases. 
To analyse this behavior, we have evaluated the deviation 
from a smooth behaviour for the dimensionless function $T\rho_N (T, \bar{s})$ 
for $7 \leq N \leq 12$ using different values of $\Delta T$ ranging from $0.005$ to $0.1$.
For any function $f(T)$, a simple measure of this deviation is the local mean square curvature
\begin{align*}
    \kappa^2_f(\Delta T)=\left\langle \left( \frac{f(T+\Delta T)-2f(T)+f(T-\Delta T)}{(\Delta T)^2} \right)^2 \right\rangle_T\,,
\end{align*}
where $\langle \cdot \rangle_T$ indicates an average over the available temperature range.
If the curve $f(T)$ is smooth, then $\kappa^2_f(\Delta T)$ does not depend on $\Delta T$
for small $\Delta T$. In contrast, \cref{fig:density}(b) shows that $\kappa^2_{T\rho_N}(\Delta T)$
blows up at small $\Delta T$. Furthermore, we see that this divergence occurs
for smaller $\Delta T$ as $N$ increases. This observation confirms 
that for a fixed $\Delta T$, larger sizes $N$ correspond to smoother transition densities.

%%%%%%%%%%%%%%%%%%%%%%%%%%%%%%%%%%%%%%%%%%%%%%%%%%%
%%%%%%%%%%%%%%%%%%%%%%%%%%%%%%%%%%%%%%%%%%%%%%%%%%%
\section{Conclusion}
%%%%%%%%%%%%%%%%%%%%%%%%%%%%%%%%%%%%%%%%%%%%%%%%%%%
%%%%%%%%%%%%%%%%%%%%%%%%%%%%%%%%%%%%%%%%%%%%%%%%%%%

As a summary, we have used dynamic programming
to compute the optimal state-dependent macroscopic field
that drives a few-particle cluster up to a desired target shape in minimum time.
The optimal policy for the field exhibits transitions
as the temperature is varied. In addition, this policy 
presents some degeneracy that is mainly controlled by symmetries
of the initial state  and of the target state.
As the size of the cluster increases, 
the number of states that are degenerate blows up exponentially
but their fraction relative to the total number of states vanishes.
Moreover, a continuum limit is found to emerge 
for the  density of temperature transitions.

Our numerical results on the asymptotic behavior (at large $N$) 
of degeneracy and temperature transitions opens novel questions for theoretical
investigations of optimal policies in
systems subject to thermal fluctuations.

Furthermore, some of our results---such as temperature transitions
or degenerate states---could be directly observable in experiments.
Indeed, as discussed in Ref.~\cite{Boccardo2022a}, electromigration
forces at the surface of metals are too small to allow one to control the shape of
few-atoms monolayer clusters.
However for colloids, where edge diffusion can be observed~\cite{Hubartt2015},
$J$ can be as small as a few $k_B T$~\cite{Nozawa2018}
using depletion interactions. Furthermore, 
as discussed in Ref.~\cite{Boccardo2022a},
thermophoretic forces~\cite{Helden2015,Braibanti2008,Wrger2010}
for polystyrene beads with a radius  2.5 \si{\micro\meter}
can lead to $F_0 \approx 10 \, k_BT/ \si{\micro\meter}$~\cite{Helden2015}. 
Thus, $F_0a/J$ can be of the order of one,
and control of few-particles clusters should therefore
be possible in experiments.
We therefore hope that experimental investigations
of the control of few-particle cluster with 
macroscopic fields will be attempted in the near future.

% \clearpage
%%%%%%%%%%%%%%%%%%%%%%%%%%%%%%%%%%%%%%%%%%%%%%%%%%%%%%%%%%%%%%%%%%%%%%
%%%%%%%%%%%%%%%%%%%%%%%%%%%%%%%%%%%%%%%%%%%%%%%%%%%%%%%%%%%%%%%%%%%%%%
%%%%%%%%%%%%%%%%%%%%%%%%%%%%%%%%%%%%%%%%%%%%%%%%%%%%%%%%%%%%%%%%%%%%%%
\appendix
%%%%%%%%%%%%%%%%%%%%%%%%%%%%%%%%%%%%%%%%%%%%%%%%%%%%%%%%%%%%%%%%%%%%%%
%%%%%%%%%%%%%%%%%%%%%%%%%%%%%%%%%%%%%%%%%%%%%%%%%%%%%%%%%%%%%%%%%%%%%%
%%%%%%%%%%%%%%%%%%%%%%%%%%%%%%%%%%%%%%%%%%%%%%%%%%%%%%%%%%%%%%%%%%%%%%
%%%%%%%%%%%%%%%%%%%%%%%%%%%%%%%%%%%%%%%%%%%%%%%%%%%%%%%%%%%%%%%%%%%%%%
%%%%%%%%%%%%%%%%%%%%%%%%%%%%%%%%%%%%%%%%%%%%%%%%%%%%%%%%%%%%%%%%%%%%%%

\section{Additional dynamical graphs}

\begin{figure}[ht]
	\includegraphics[width=.5\linewidth]{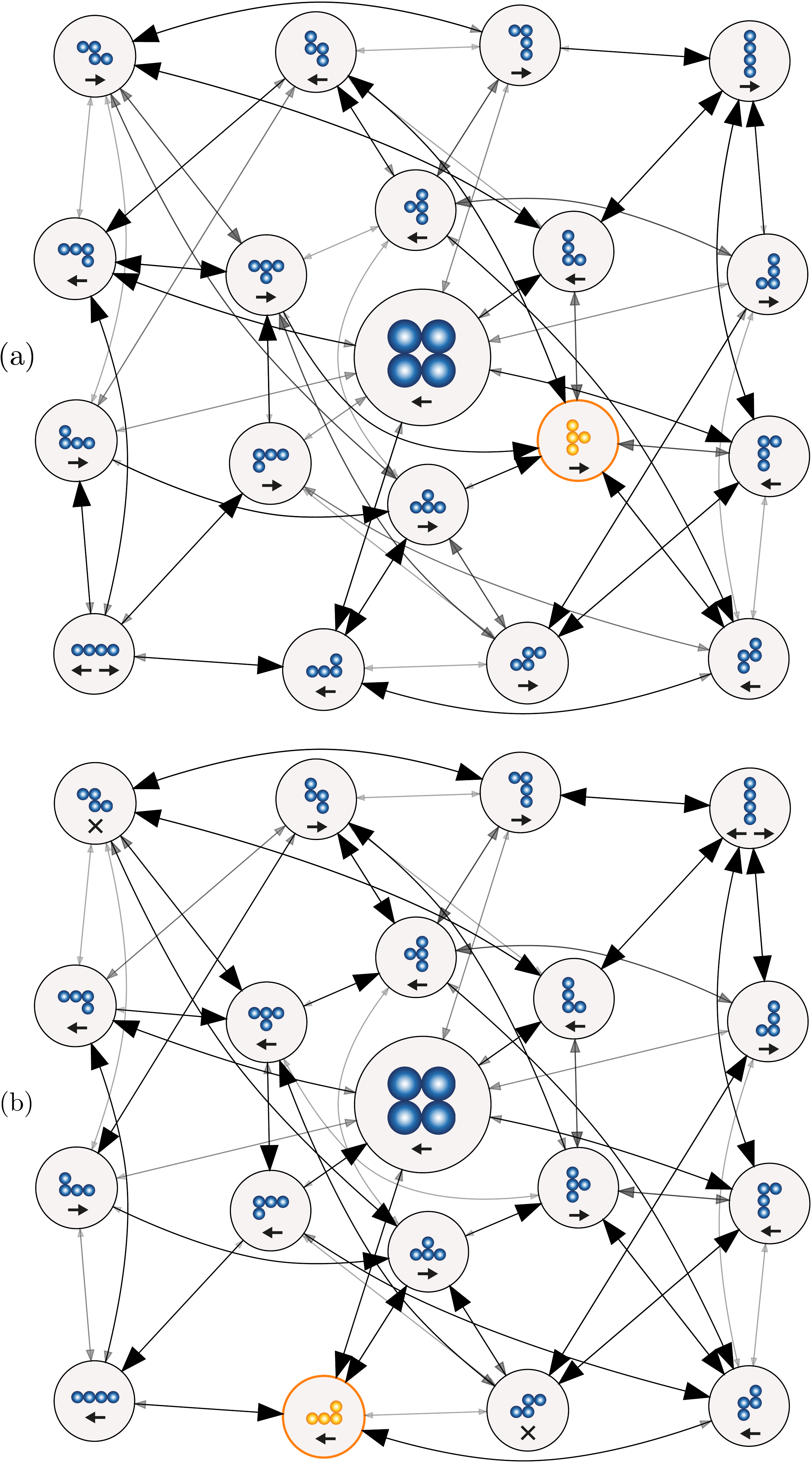}
	\caption{Dynamical graph of two tetramer clusters at $T=0.24$, with the optimal policies to reach the two orange targets. The target in (a) belongs to the $\CMcal{H}$ symmetry class, while the one in (b) belongs to the $\CMcal{N}$ symmetry class. In both cases, there is one state (bottom left for (a) and top right for (b)) that shows a degenerate optimal action.
	}
	\label{fig:graph_tetramer_nontrivial_app}
\end{figure}

We report here some additional dynamical graphs. In \cref{fig:graph_tetramer_nontrivial_app} we show two cases of tetramer targets that do not have the $x\rightarrow-x$ symmetry, but that still exhibit a degenerate state. All the 10 cases of degeneracy due to \q{hidden} symmetries of the graph in tetramers can be obtained by 90 degree rotation, left-right, or up-down symmetry transformations of the three targets shown in \cref{fig:graph_tetramer_nontrivial,fig:graph_tetramer_nontrivial_app}.

\begin{figure}[ht]
	\includegraphics[width=.5\linewidth]{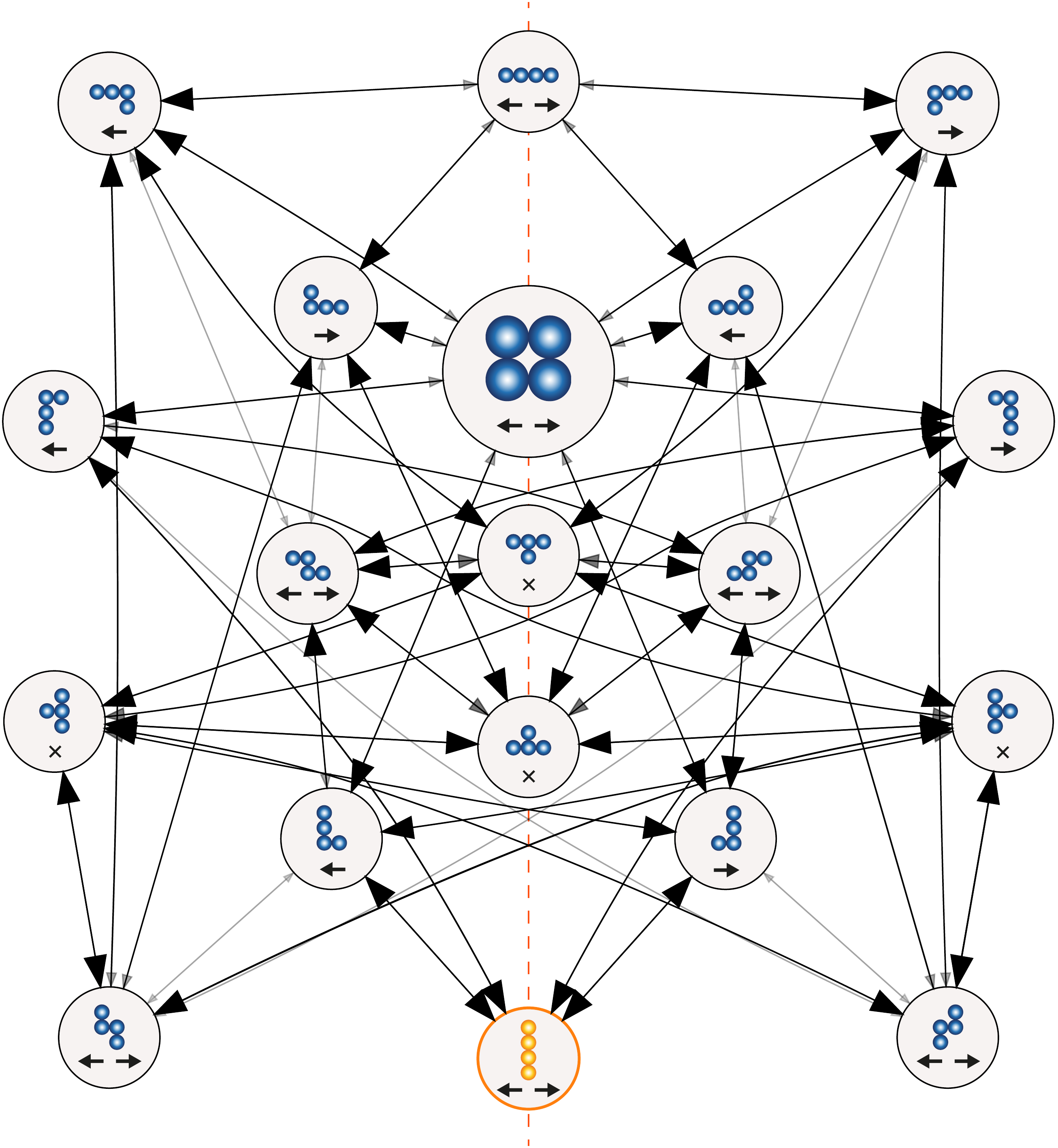}
	\caption{Dynamical graph of a tetramer cluster at $T=0.24$, with the optimal policies to reach the orange target. This target belongs to the $\CMcal{V}$ symmetry class, hence, all states that are invariant with respect to the $x\rightarrow-x$ transformation (shown here on the vertical axis of the graph) are expected to be either degenerate or with a zero-force optimal action (indicated with a cross).
	}
	\label{fig:graph_tetramer_symmetric}
\end{figure}

In \cref{fig:graph_tetramer_symmetric} we report a dynamical graph showing that the states that obey the symmetry rules of degeneracy are also compatible with a zero-force optimal action.

\bibliography{references}

\end{document}